\documentclass[traditabstract]{aa}

\usepackage{graphicx}
\usepackage[svgnames]{xcolor}
\usepackage[pdftex,colorlinks,citecolor=DarkBlue]{hyperref}

\usepackage[varg]{txfonts}

\usepackage{natbib}
\bibpunct{(}{)}{;}{a}{}{,} 

\newcommand{\teff}{\mathrm{T_{eff}}}
\newcommand{\mh}{{\mathrm{M}_{\rm HBMM}}}
\newcommand{\msol}{{\mathrm{M}_\odot}}
\newcommand{\mjup}{{\mathrm{M_{Jup}}}}
\newcommand{\rjup}{{\mathrm{R_{Jup}}}}

\newcommand{\mdyn}{{\mathrm{M_{dyn}}}}

\def\msol{{\rm M}_\odot}

\begin{document}

   \title{Impact of a new H/He equation of state on the evolution of massive brown dwarfs. New determination of the hydrogen burning limit}

  \titlerunning{Impact of a new H/He equation of state on the evolution of massive brown dwarfs}

   \author{Gilles Chabrier \inst{\ref{inst1},\ref{inst2}}
      \and Isabelle Baraffe \inst{\ref{inst2},\ref{inst1}}
              \and Mark Phillips \inst{\ref{inst3},\ref{inst2}} 
              \and Florian Debras \inst{\ref{inst4}}
     }

   \institute{Ecole normale sup\'erieure de Lyon, CRAL,  CNRS UMR 5574, 69364, Lyon Cedex 07, France
      \label{inst1}
      \and
     Physics \& Astronomy Dpt, University of Exeter, Exeter, EX4 4QL, UK
      \label{inst2}
      \and
      Institute for Astronomy, University of Hawaii at Manoa, Honolulu, HI 96822, USA
       \label{inst3}
      \and
       IRAP, Universit\'e de Toulouse, CNRS UMR 5277, UPS, Toulouse, France
      \label{inst4}
      }

   \date{Received ---; accepted ---}

       \abstract%
      {We have explored the impact of the latest equation of state (EOS) for dense hydrogen-helium mixtures (Chabrier \& Debras 2021), which takes into account the interactions between hydrogen and
helium species, upon the evolution of very low mass stars and brown dwarfs (BD).
These interactions modify the thermodynamic properties of the H/He mixture, notably the entropy, a quantity of prime importance for these fully convective bodies, but also the onset and the development of degeneracy throughout the body. This translates into a faster cooling rate, i.e. cooler isentropes for a given mass and age, and thus larger brown dwarf masses and smaller radii for given effective temperature and luminosity
 than the models based on previous EOSs. This means that objects of a given mass and age, in the range $M\lesssim 0.1\,\msol$, $\tau\gtrsim 10^8$ yr, will have cooler effective temperatures and fainter luminosities. 
Confronting these new models with several observationally determined BD dynamical masses, we show that this improves the agreement between evolutionary models and observations and resolves at least part of the 
observed discrepancies between the properties of dynamical mass determinations and evolutionary models.
A noticeable consequence of this improvement of the dense H/He EOS is that it yields a larger H-burning minimum mass, now found to be $0.075\,\msol$ ($78.5\,\mjup$) with the ATMO atmosphere models for
solar metallicity. These updated brown dwarf models are made publicly available.
}


   \maketitle


\section{Introduction}

Tremendous progress has been accomplished along the years in the theoretical description of brown dwarfs (BD), enabling us to better understand their fundamental properties and cooling histories. The most recent progresses include a more complete description of their atmosphere,  thus of their spectral energy distribution (e.g. Morley et al. 2012, 2014, Tremblin et al. 2015, 2016, 2017), and of their interior, notably concerning the
equation of state (EOS) of dense hydrogen and helium (Chabrier et al. 2019, CMS19). A new generation of BD evolutionary models has been derived recently, that incorporates both new (so-called 'ATMO') atmosphere models
and the new CMS19 H/He EOS (Phillips et al. 2020). As shown in these models, a noticeable impact  of this new EOS is to yield denser and cooler, thus more degenerate objects at the same mass than
the ones computed with the Saumon-Chabrier-vanHorn (SCvH) EOS (Saumon et al. 1995). This yields slightly faster
cooling rates, thus cooler temperatures and lower brightness at a given age (Phillips et al. 2020).

Yet, a puzzling issue has emerged within the recent years, thanks to the determination of the dynamical masses of several 'massive' T dwarfs (Cheetham et al. 2018, Dieterich et al. 2018, Dupuy et al. 2019, Bowler et al. 2018, Sahlmann et al. 2020, Brandt et al. 2019). These observations reveal dynamical masses significantly larger than
the theoretical predictions for the determined effective temperature and age. A common feature of all these objects is their relatively high mass ($\sim60$-75 $\mjup$), near the hydrogen burning minimum mass (HBMM), similar spectral type (late T) and cool effective temperatures ($\teff\lesssim 1200$ K), which suggests ages in the range $\sim 5$-10 Gyr, at the very least $>1$ Gyr. A noticeable exemple of this
puzzling issue is the recent dynamical mass determination of Gliese 229 B, with a larger mass than estimated previously (Brandt et al. 2020). For all these late, massive T dwarfs, all existing  BD evolutionary models generally underpredict the mass for the nominal ages and temperatures, suggesting too small cooling rates for these objects. 
This trend is confirmed by the recent thorough analysis of BD companion dynamical masses using Hipparcos and Gaia EDR3 data and age determinations based on activity-rotation-age calibrations by Brandt et al. (2021). For old ($\gtrsim$5 Gyr) and high-mass  ($\gtrsim 60\,\mjup$) BDs,
the models overpredict luminosities for the measured mass and age, or equivalently underestimate (resp. overestimate) the mass (resp. the age) for the correct age (resp. mass).
As noticed by Dieterich et al. (2018), concerning the Eps Indi B-C system, models cannot make such massive objects reach such a cool $\teff$ within the age of the Galaxy.
Due to the degeneracy between mass and age in the substellar domain, there are no reliable age indicators for isolated BDs, and thus no robust constraint can be derived regarding substellar cooling rates and evolutionary models. In contrast, the aforementioned BD companions to higher mass stars with masses close to the stellar-substellar limit allow us to test a boundary value of the theory of substellar structure and evolution. 

A related diagnostic of these observations is the suggestion that the HBMM, identified as a minimum in the radius-effective temperature and radius-luminosity relation, is larger and lies at a $\sim 400$ K larger effective temperature than predicted by all current models (Dieterich et al. 2014).

In this paper, we show that the most recent improvements in EOS calculations of dense hydrogen/helium mixtures contribute to resolve these issues in BD cooling theory.
From a more general perspective, it should be mentioned that the aforementioned recent analysis of Brandt et al. (2021) suggests another, different issue for young ($<$1 Gyr) and low-mass ($\lesssim 40\,\mjup$) BDs. In that case, the trend is the opposite: models  underpredict luminosities for given mass and age, or equivalently overestimate (resp. underestimate) the mass (resp. the age) for the correct age (resp. mass). For objects in-between, mass $\sim$40-70 $\mjup$, age $\sim$1-5 Gyr, models agree well
($<1\sigma$) with observations (see Table 10 of Brandt et al. (2021)). This points to two different issues in BD cooling theory in the extreme mass and age limits.
    In the present paper, we will focus on the first issue, for which we show that part of the solution lies in the EOS. In contrast, the second problem stems more likely from remaining uncertainties in BD atmosphere models.

\section{Impact of the new equations of state on the internal structure, cooling history and hydrogen-burning limit}
\label{theory}

\setlength{\unitlength}{1cm}
\begin{figure}
{\includegraphics[width=8.5cm]{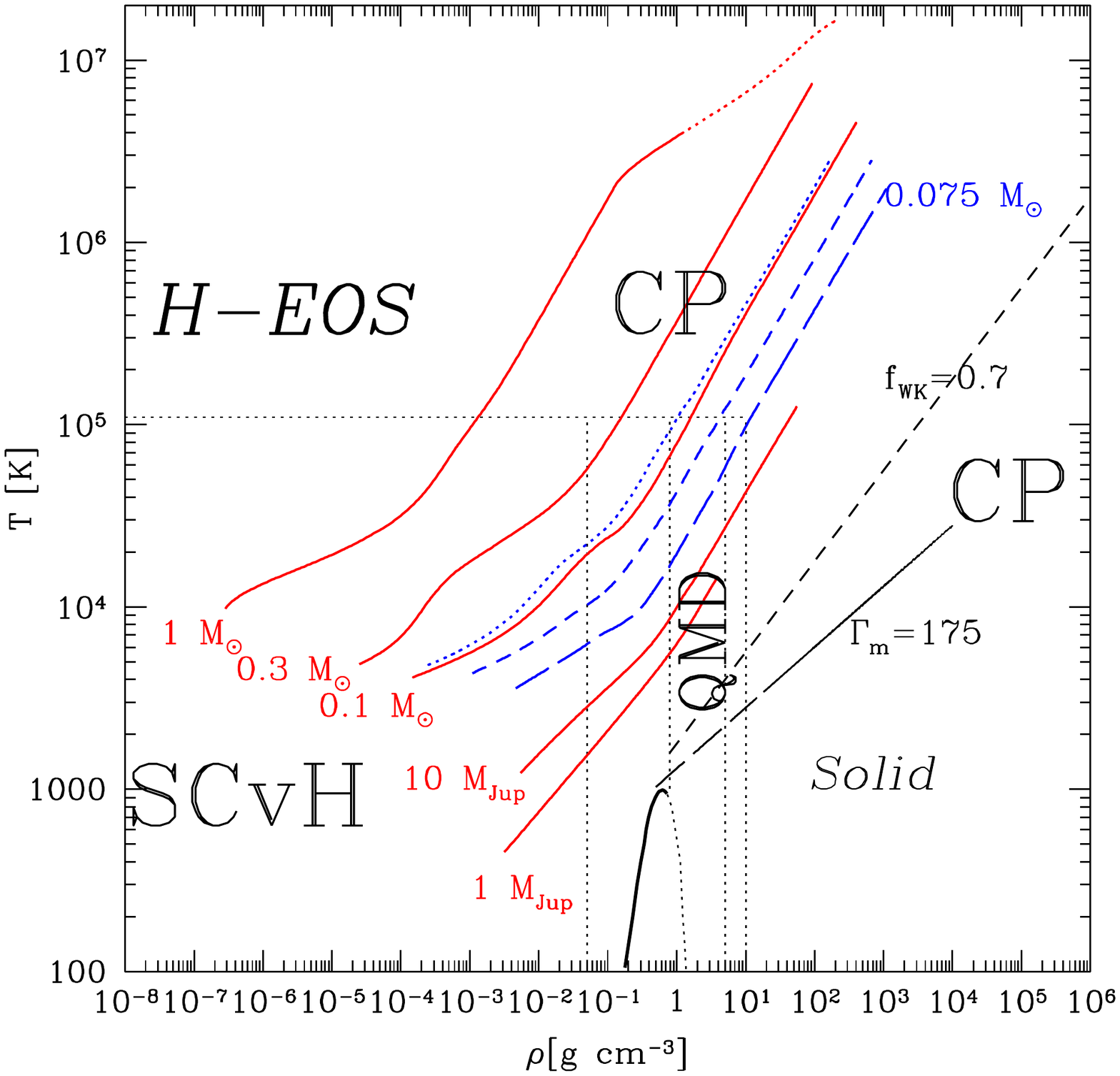}  }
\vspace{-2.cm}
\caption{Interior $T$-$\rho$ profiles in the hydrogen phase diagram for different astrophysical bodies at about 5 Gyr, as labelled in the figure (red lines) and  for a 0.075 $\msol$ object at 10$^8$, $ 8\times$10$^8$ and $8\times$10$^{9}$ yr
(blue dotted, short-dashed and long-dashed lines, respectively). The SCvH, CP and QMD labels stand for Saumon, Chabrier \& vanHorn (1995) EOS, Chabrier \& Potekhin (1998) EOS and QMD simulations (see Chabrier et al. (2019) and Chabrier \& Debras (2021) for further details on the line symbols).}
\label{diag}
\end{figure}

\setlength{\unitlength}{1cm}
\begin{figure*}
{\includegraphics[width=9.5cm]{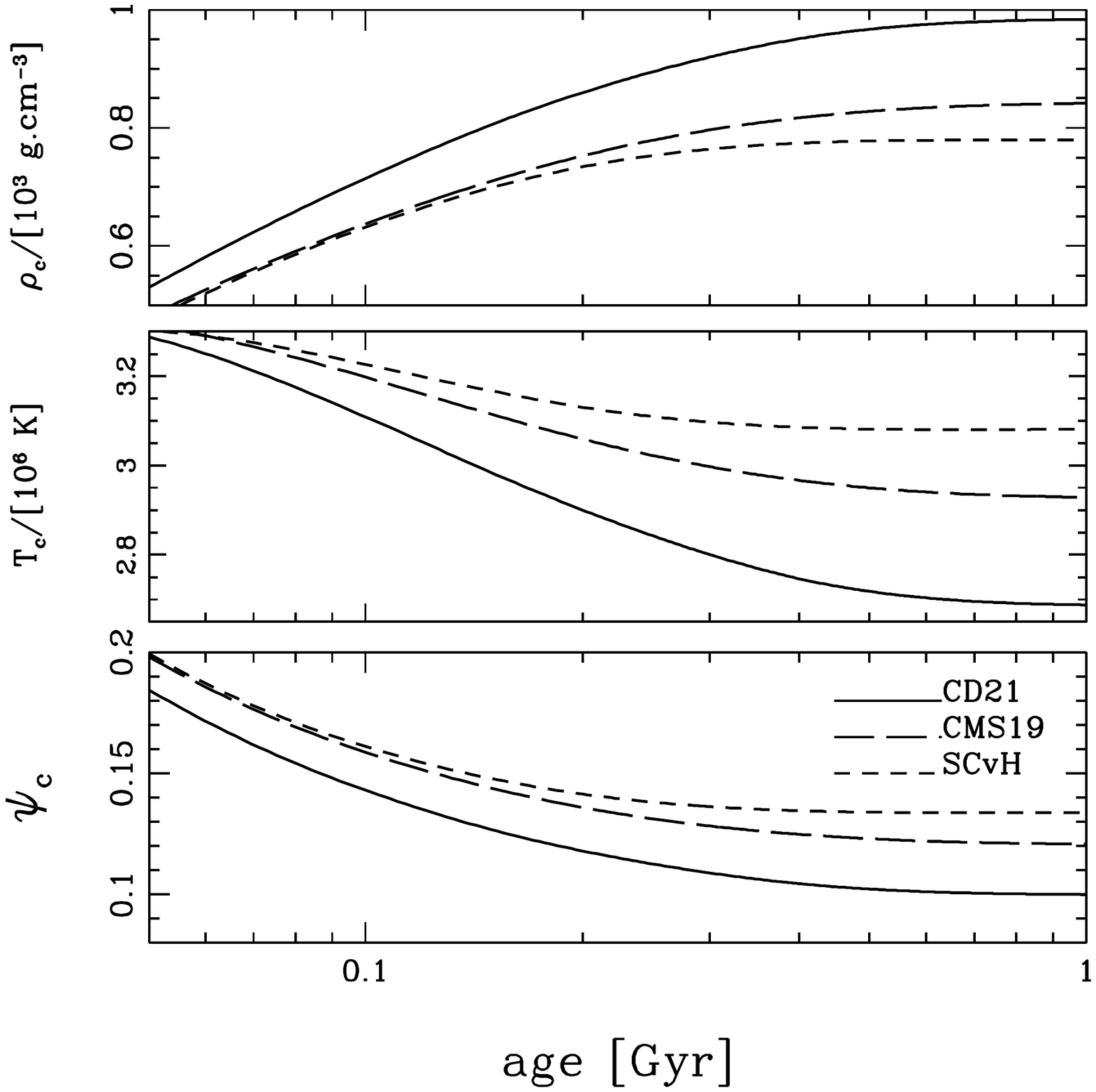}}  
\hspace{-1.cm}
{\includegraphics[width=9.5cm]{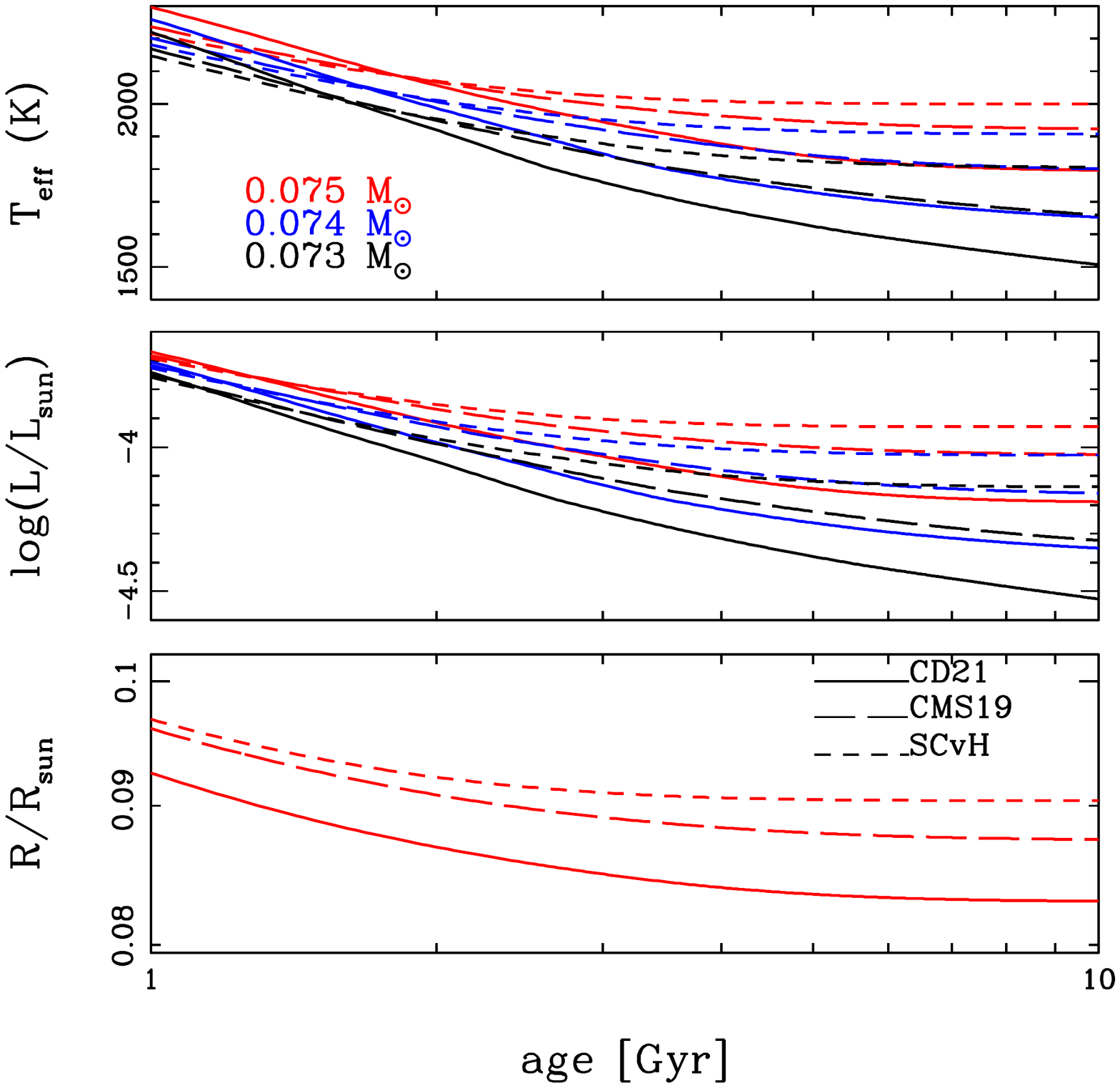}}  
\vspace{-2.cm}
\caption{Left: Evolution of the central density, temperature and degeneracy parameter for a 0.075 $\msol$ object calculated with 3 different EOS, namely CD2021 (solid line), CMS2019 (long-dashed line)
and SCvH (dotted line). Right: Evolution of the effective temperature and luminosity for 0.075 (red), 0.074 (blue) and 0.073 (black) $\msol$   and radius  for 0.075  $\msol$, with the same 3 EOS.}
\label{evol}
\end{figure*}

 \begin{table*}
         \begin{center}
            \begin{tabular}{lcccc}
               \hline\hline
               EOS & $\mh/M_\odot$  & ${\rm T_{{eff}}}_{\rm HBL}$ [K] &  $\log(L/L_\odot)_{\rm HBL}$ & $R_{\rm HBL}/R_\odot$ \\
               \hline
               SCvH+COND (Baraffe et al. 2003) &  0.072   & 1560     & -4.47 & 0.081  \\
               SCvH+ATMO &  0.073   & 1807     & -4.14 & 0.087  \\
               CMS'19+ATMO (Phillips et al. 2020) &  0.074 & 1800    &   -4.16 &  0.085 \\
               CD'21+ATMO (present)  & 0.075  &  1800   & -4.19   & 0.083\\
                              \hline
            \end{tabular}
         \end{center}
         \caption{Mass, effective temperature, luminosity and radius characteristic of the H-burning limit ($t$=10 Gyr) with the SCvH (1995), Chabrier et al. (2019) and Chabrier \& Debras (2021) EOSs, all calculated with the ATMO atmosphere models. For the sake of comparison, the results with the COND atmosphere models are also presented. }
\label{table_HBL}
      \end{table*}

Figure \ref{diag} displays temperature-density profiles of various astrophysical bodies, from 1 $\msol$ to 1 $\mjup$ for an age of about 5 Gyr, in the phase diagram of hydrogen, for the
$T$-$\rho$ range covered by the CMS19 and Chabrier \& Debras (2021, CD21) EOSs. This diagram is similar to the ones portrayed in these papers for H and He and we refer the reader to these
papers for more details. The key issue here is that, as illustrated by  the central part of the diagram labelled QMD (for Quantum Molecular Dynamics), all BD cooling tracks enter the domain where interactions between H and He species
can no longer be ignored, as assumed in the so-called ideal (or 'additive') volume law approximation, used in CMS19 and Saumon et al. (1995, SCvH) (see Fig. 1 of CD21 for further details). Recently, Chabrier \& Debras (2021) took into account the
impact of these interactions on the thermodynamic properties of the H/He mixture by incorporating in the EOS the QMD calculations performed by Militzer \& Hubbard (2013). The figure also
displays the cooling sequence of a 0.075 $\msol$ object at 10$^8$, $\sim$10$^9$ and $\sim$10$^{10}$ yr, respectively. As seen in the figure (see also the Conclusion of CMS19), essentially all objects below
$0.1\,\msol$ older than about $\sim 0.1$ Gyr will enter the domain where H-He interactions cannot be ignored and will affect to some degree their mechanical and thermal properties, thus their
structure and evolution. It is the aim of the present paper to examine in detail this impact.
      
      \setlength{\unitlength}{1cm}
\begin{figure*}
{\includegraphics[width=7.cm]{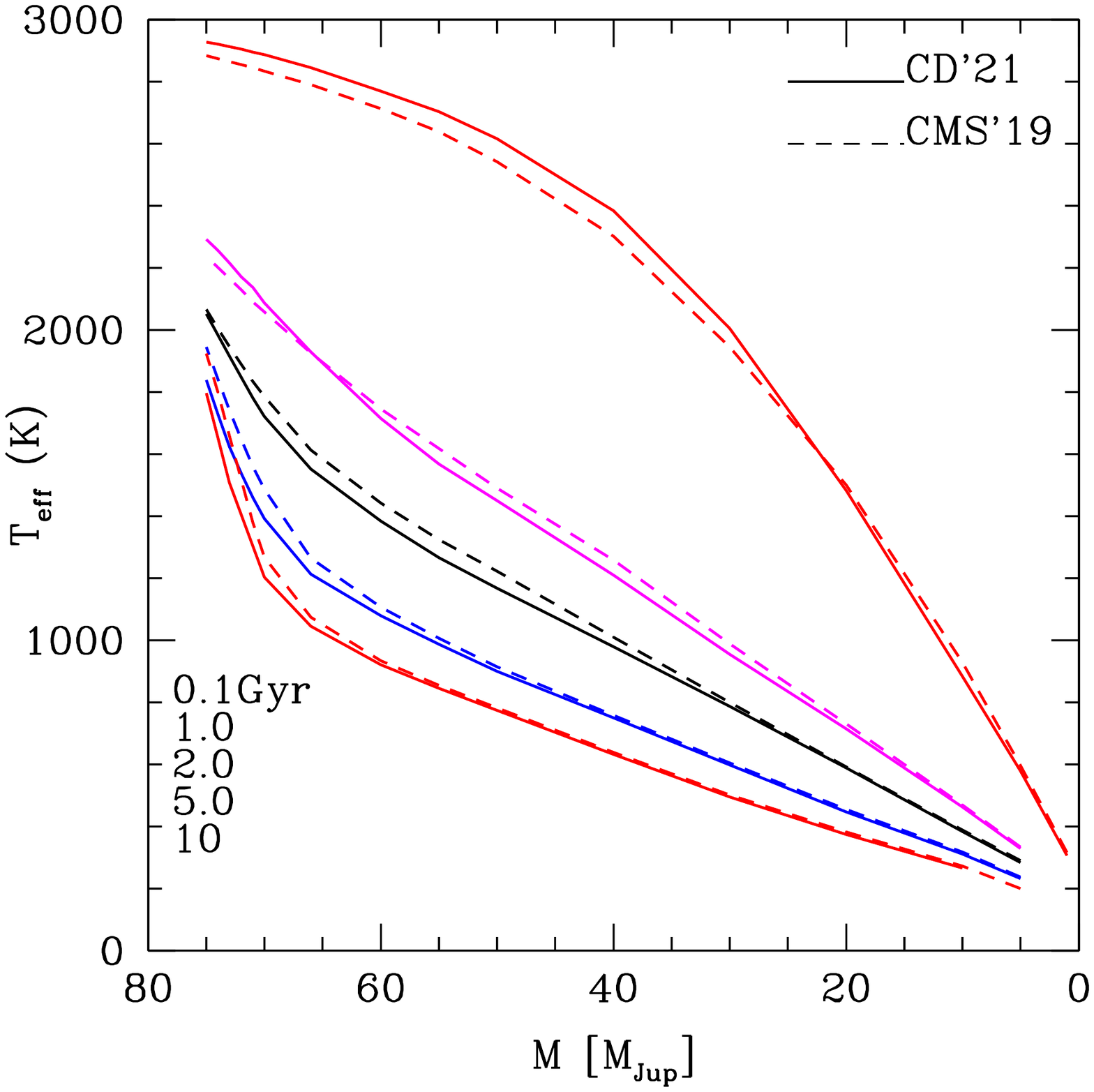}}  
\hspace{-1.cm}
{\includegraphics[width=7.cm]{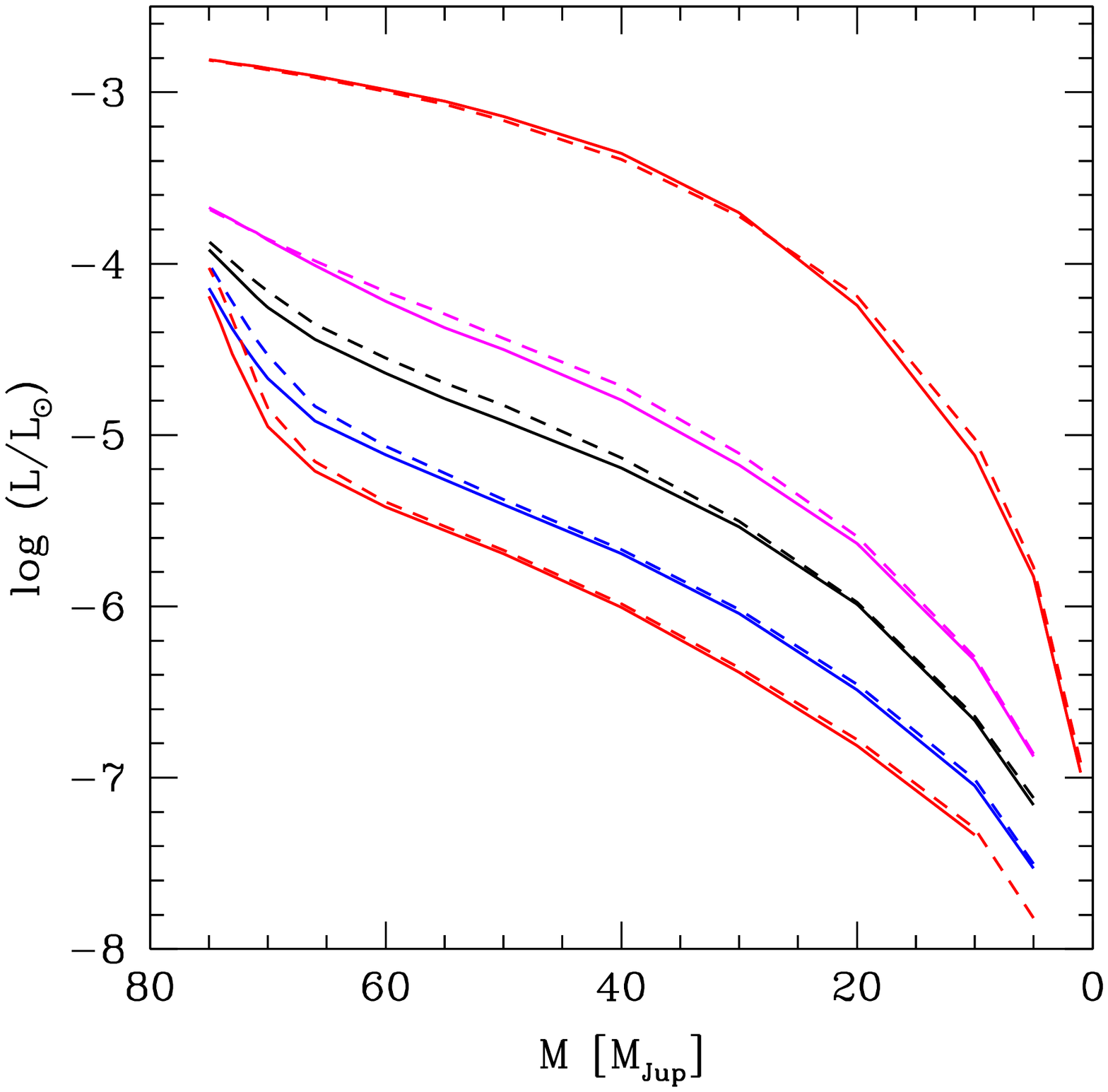}}  
\hspace{-1.cm}
{\includegraphics[width=7.cm]{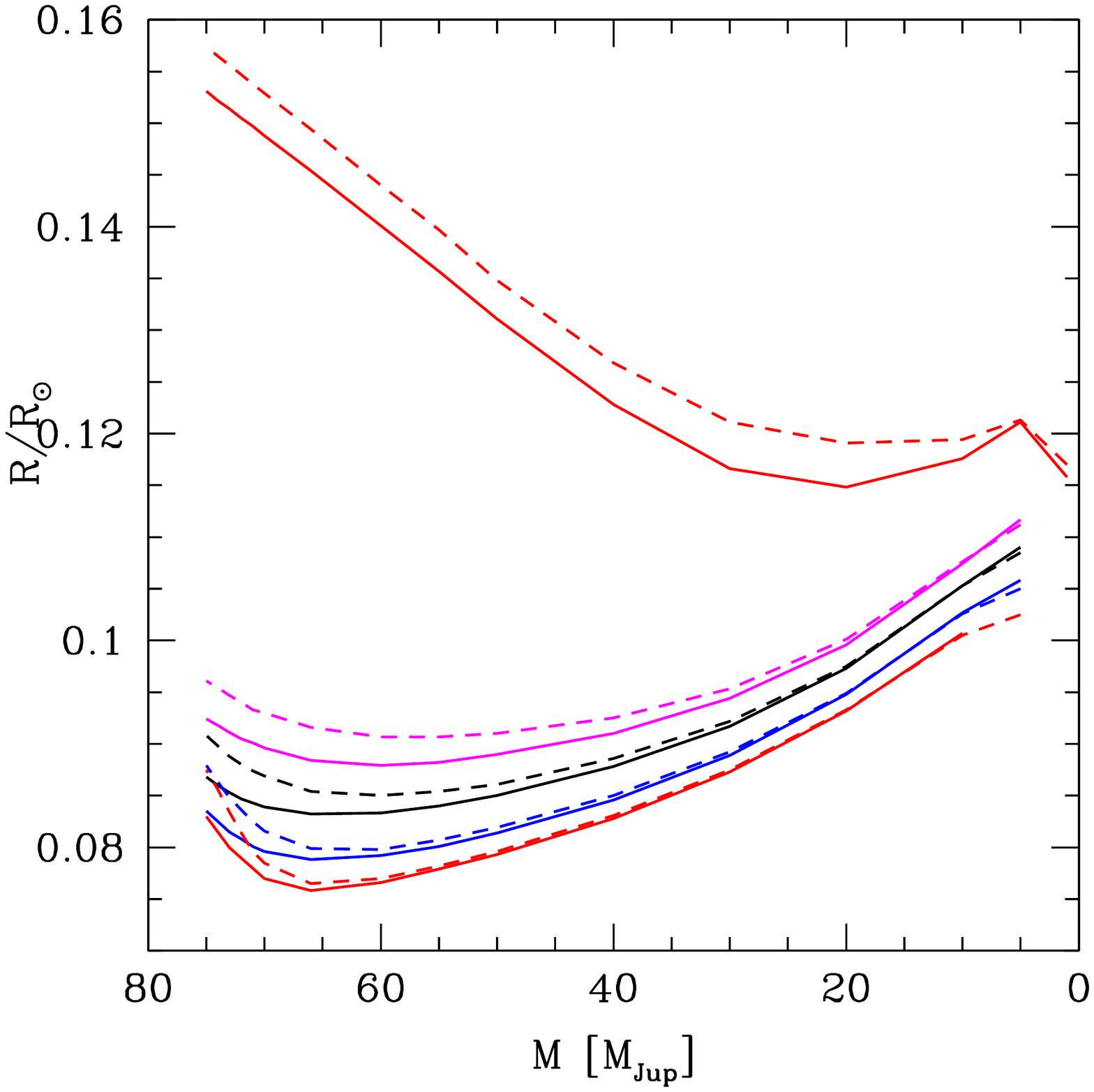}}  
\vspace{-2cm}
\caption{Effective temperature, luminosity and radius for $M\le0.075\,\msol$ for 0.1, 1, 2, 5 and 10 Gyr (from top to bottom), respectively, for the same atmosphere models and helium and heavy element compositions but 2 different EOSs, namely Chabrier \& Debras (2021) (present) and Chabrier et al. (2019) (Phillips et al. 2020).}
\label{evol_m}
\end{figure*}

The left panel of Fig. \ref{evol} displays the evolution of the central density, $\rho_c$, temperature,  $T_c$, and degeneracy parameter, $\psi_c=(T_c/T_F)\approx 3.314\times10^{-6}T_c(\mu_e/\rho)^{2/3}$,where $T_F$ and $\mu_e$ denote the electron Fermi temperature and mean molecular weight, respectively,  of a 0.075 $\msol$ object (see Chabrier \& Baraffe 2000). The evolution is calculated with
our recent  so-called ATMO atmosphere models and updated solar abundances (Caffau et al. 2011) for a global helium and heavy element abundance $Y=0.275, Z_\odot=0.017$, respectively (Phillips et al. 2020), but with 3 different H/He EOS, namely Saumon et al. (1995, SCvH), Chabrier et al. (2019, CMS19) and Chabrier \& Debras (2021, CD21). The differences between these different EOS, and the improvements in the treatment of the 
H-He interactions, are described in detail in these papers. As already mentioned in Phillips et al. (2020) and Chabrier \& Debras (2021) and clearly seen in the figure, the new EOSs yield denser and cooler structures for a given object, thus more correlated and degenerate (lower $\psi$) interiors. This in turn increases the cooling rate of the object. This is illustrated in the right panel of Fig. \ref{evol}, which portrays the late evolution of the effective temperature, luminosity and radius of a 0.073 (black), 0.074 (blue) and 0.075 (red) $\msol$ object for the same 3 EOSs. While these 3 quantities, notably $\teff$ and $L$, become constant after $\sim 5\times 10^8$ yr at $0.073\,\msol$ with the SCvH EOS, indicating the stellar-substellar boundary,   they keep decreasing with the two other EOSs, this limit occuring at $0.074\,\msol$ with the CMS19 EOS and at $0.075\,\msol$ with the CD21 EOS. 
Table \ref{table_HBL} displays the characteristic of the steady H-burning limit (HBL) obtained with the 3 EOSs. We define the HBL as the limit below which nuclear equilibrium ($L_{\rm nuc}=L_\star$) will never be reached,  thus below which cooling and gravitational contraction will last for ever. This corresponds to the physical limit of the stellar main sequence, i.e.
the stellar-substellar boundary. Any object below this limit, i.e. cooler and fainter than the corresponding effective temperature and luminosity (see Table 1) will be a brown dwarf, whatever its age. Note, however, that the opposite is not true: objects hotter and brighter than these limits can be either stars or BDs, depending on their age. The H-burning minimum mass (HBMM),
i.e. the minimum mass to sustain hydrogen fusion, that does not depend on the age, is  found to be 0.075 $\msol$ ($\sim78.5\,\mjup$) with the most recent CD21 EOS. For sake of comparison, we also give in this table the
characteristics of the HBL obtained in Baraffe et al. (2003). These models use the SCvH EOS, the same helium and heavy element abundances as mentioned above but the so-called COND model atmospheres. These comparisons enable us to disentangle the impact of the atmosphere and EOS models, respectively, upon the cooling properties of an object at the H-burning limit. 
The atmosphere models have been greatly improved between the COND and ATMO models. Line opacities, in particular, were missing in the former ones, yielding a much faster cooling rate and thus explaining the cooler and fainter limits at the HBL. But the EOS also bears some impact on the HBL. 
As seen in Table \ref{table_HBL}, the larger cooling rate with the new EOS translates into a larger HBMM with a smaller radius than in the previous calculations. On the other hand, the luminosities at the HBL remain barely affected ($\lesssim 10\%$) and the effective temperature remains essentially the same. This is explained by the fact that the threshold for hydrogen fusion at the center of the star occurs at a fixed temperature and thus, for the same interior-atmosphere (T$_c$-$\teff$) boundary condition, at the same effective temperature. In contrast, the nuclear energy rate, thus the luminosity depends on the mass and then slightly differs between the 3 different HBMM's.

On the other hand, Dieterich et al. (2014) have determined the
effective temperature and bolometric fluxes of 63 objects
ranging in spectral type from M6V to L4, bracketing the stellar/substellar boundary, by comparing observed optical and infrared photometric colors on 9 bands with synthetic ones, derived from the so-called BT-Settl model atmospheres (Allard et al. 2012, 2013), used in the Baraffe et al. (2015) evolutionary calculations. The optimisation procedure for determining effective temperatures also  indicates which model spectrum in the BT-Settl grid provides the overall best fit to the observed photometry. Once the effective temperatures and bolometric fluxes of the objects were determined by this procedure, the radii of the objects with known trigonometric parallax were determined from the Stefan-Boltzmann law (see Dieterich et al. (2014) for details).
As the differences between stellar and substellar objects become more pronounced at ages $>1$ Gyr, objects with known youth signatures were rejected from the sample.
Based on this analysis, Dieterich et al. (2014) find evidence for a local minimum in the radius-temperature and
radius-luminosity trends,  that indicates the vicinity of the stellar main sequence - brown dwarf sequence boundary (see e.g. Burrows et al. 1997, Chabrier \& Baraffe 2000, Chabrier et al. 2009), 
at ${\rm T_{{eff}}}_{\rm HBL} \simeq 2075$ K, $\log(L/L_\odot)_{\rm HBL}\simeq -3.9$ and $R/R_\odot \simeq 0.086$.
The inversion of the radius trend occurs near the location of the L2.5 dwarf 2MASS J0523-1403 ($M=67.54\pm12.79\,\mjup$, Sp=L2.5 (Filippazzo et al. 2015)). Although at first glance this seems to point to a disagreement with
the models in the determination of the HBL, it must be kept in mind that the age of the objects in the Dieterich et al. (2014) sample, although supposedly larger than
1 Gyr, is unknown. This inevitably translates into some uncertainty in the comparison between observations and evolutionary calculations. As seen in the right panel of Fig. \ref{evol}, however, the HBL is only reached at much older ages, about 10 Gyr with the new EOS.
Therefore, most of the objects identified in Dieterich et al. (2014), which lie at less than 25 pc and belong dominantly to the young disk, are too young to yield
a proper determination of the HBL. On the other hand, as seen in figure \ref{evol_m},  which portrays the $\teff$-$M$, $L$-$M$ and $R$-$M$ relations below the HBMM for 5 isochrones, an average age of about 2 Gyr for the observed sample near the $R$-$\teff$ and $R$-$L$ minimum location  yields an excellent agreement between the observational determinations and the new models. As highlighted in the rightmost panel of this figure, the minimum in the radius-mass relation does not occur exactly at the HBMM but at about 65 $\mjup$ (see, e.g., Fig. 1 of Chabrier et al. 2009),
consistent with the aforementioned BD 2MASS J0523-1403 determination,
and is more and more pronounced as objects get older. We  see from the 3 panels that the observational identifications of the minimum radius in the sample of Dieterich et al. (2014) are consistent with objects of mass $\sim 0.065$-$0.075\,\msol$ and age $\sim1$-2 Gyr. We also see that, whereas the newest EOS affects only modestly the effective
temperature and the luminosity, for a given mass and age, compared with the CMS19 EOS used in the models of Phillips et al. (2020), it affects more severely the radius, notably for the most massive BDs. This reflects the different rate at which degeneracy progresses throughout the body, starting at about $\sim 0.1$ Gyr, as seen in the left panel of Fig. \ref{evol}. The Phillips et al. (2020) models have been updated accordingly and are available online (see at the end of the paper). 

 It is important to mention at this stage that the
 procedure used by Dieterich et al. (2014) remains model atmosphere dependent, notably in a temperature domain characterised by cloud condensation and sedimentation in the models. Even though the BT-Settl atmosphere models, used by Dieterich et al. (2014), have been shown to provide a remarkable agreement with observed M and L spectra, alternative 'cloudless' models have been proposed, which are very successfull in reproducing BD atmosphere spectra  (Tremblin et al. 2015, 2016, 2017). 'Cloudless' in
these models does not mean that clouds do not form, but rather that they are not responsible for shaping the spectral evolution of brown dwarfs. Instead, the (massive) BD spectral evolution is due to a thermo-compositional instability 
triggered by the chemical conversion $CO\rightarrow CH_4$ in their atmospheres (Tremblin et al. 2019). 

Concerning the domain of  very-low mass stars (VLM, $M\le 0.4\,\msol$) on the main sequence (age $\gtrsim 1$ Gyr), the impact of the new EOS (CD21) compared with the SCvH one used in the Baraffe et al.  (2015, BHAC) models is quite modest and thus will not be displayed here.

\section{Comparison with the data}

As mentioned in the previous section, near-infrared spectroscopic and imaging surveys have uncovered a population of short-period,
spectral binaries composed of low-mass stars and brown dwarfs, allowing a precise dynamical determination of the mass of these latter objects.
Activity and kinematic constraints for the age of the primary of these massive T-dwarfs lead to a range of 2-8 Gyr.
In this section, we compare the effective temperatures obtained with the COND models (Baraffe et al. 2003), the Phillips et al. (2020) models, based on the Chabrier et al. (2019) EOS, and a subset of models
based on the most recent Chabrier \& Debras (2021) EOS, with the various observations of BD dynamical systems. 

 \begin{table*}
         \begin{center}
            \begin{tabular}{lcccccc}
               \hline\hline
               Object     &  &  $M$ [$\mjup$] & age [Gyr]           &                        $\teff$ [K]          &  $\log(L/L_\odot)$ &  $R/R_\odot$ \\
               \hline \\
               HD4113C & Observations$^{1,8}$ &  66$\pm 5$         & 5$^{+1.3}_{-1.7}$ &   500-600            &  $-6.30\pm 0.22$     &                    \\
                             &  P2020            &    63                     &  5                           &  1108   &   -5.06 &      0.080                \\
                             &                         &     66                     &                             &  1149   & -4.99  &     0.080                \\
                              & {\bf present}    &    63                &  5                           &  1081   & -5.11  &    0.079                     \\
                             &                          &     66                     &                             & 1144   &  -5.02 &  0.079       \\\\
             Eps Indi Ba  & Observations$^2$ &  66.92$\pm 0.36$         & 3.5$^{+0.8}_{-1.0}$  &            &            &             \\
                                 & Observations$^5$ &  68$\pm 0.9$         &   &    1312$\pm9$          &      -4.70$\pm 0.02$      &               \\
                                           &    P2020           &   66-67                     &   3.0                       & 1324-1395 &   -4.72 -  -4.63     &        0.083       \                 \\
                                           &                    &                                     &   3.5                       & 1262-1325   & -4.81 -  -4.73   &       0.082        \                 \\
                                           &                    &                                     &   3.8                       & 1233-1293 &   -4.85 - -4.78   &       0.081        \                 \\
                                           &    {\bf present}            &     66-67         &  2.8       & 1321-1354 & -4.74 - -4.70  &     0.081          \                 \\
                                           &                &                                &  3.0             & 1291-1321 &  -4.78 -  -4.74 &     0.081          \                 \\
                                           &                &                                &  3.5             & 1241-1265 &  -4.86 - -4.83 &      0.080         \                 \\\\
                 Eps Indi Bb (or C) & Observations$^2$ &  53.25$\pm 0.29$         &  3.5$^{+0.8}_{-1.0}$ &          &        &               \\
                                               & Observations$^{5}$ &  53$\pm 0.3$         &  &    975$\pm 11$  &   -5.23$\pm 0.02$ &   \\
                                          &    P2020           &   53                     &  3.0                   & 1071   &   -5.08   &      0.084          \                 \\
                                          &                        &                          &  3.5                   & 1023 &   -5.17  &         0.083       \                 \\
                                          &                        &                          &  3.8                   & 996 &   -5.21  &         0.083       \                 \\
                                          &     {\bf present}                 &     53                     &  2.8             &  1060 & -5.10  &    0.083             \\ 
                               &                                    &                             &  3.0             &  1042 & -5.13  &    0.083             \\ 
                               &                                   &                            &  3.5             &  998 & -5.22  &    0.083         \ \\
                               &                                   &                          &  3.8             &  973 & -5.26  &     0.082            \\\\
             GL 758 B & Observations$^3$ &  42$^{+19}_{-7}$ $(>30)$     & 1-6\,({\rm older?}) &   650           &     -6.07$\pm 0.03$      &                  \\
                             &  Observations$^4$ &   $38.1^{+1.7}_{-1.5}$           & $\gtrsim 6$           &                    &     &  \\
                               &     P2020                &   37                     &  5                           &  680  &  -5.84    &    0.087            \\
                               &                                &                          &  8                           & 600   & -6.07  &    0.086               \\
                                &      {\bf present}                &   37                     &  5               & 673   & -5.86   &   0.087         \\
                              &                     &                                                  &  8               & 593   & -6.09    &   0.085         \\\\
           WISE  J0720-0846B  & Observations$^6$ &  66$\pm 4$     &  $>$ a few    &   1250$\pm$40   &  -4.82$\pm 0.07$    &               \\
                                               &     P2020                     &  66                    &  3   &   1324  &  -4.72    &    0.082            \\
                                               &     &                                                  &  4   &  1216  &  -4.88    &    0.082            \\
                                            &      {\bf present}                &   66                     &  3   & 1291   & -4.78    &  0.081          \\
                                           &                                          &                         &  4   & 1201   & -4.92    &  0.080          \\\\
        2M0805+48    & Observations$^7$ &  66$^{+5}_{-14}$     & $\ge 4$  &   &           &                            \\
       2M1059-21    & Observations$^7$ &  $67^{+4}_{-5}$     &   &   &            &                            \\
                                          &     P2020                &   66                     &  5                           &  1149  &  -4.99    &    0.080            \\
                                         &                     &                        &  10                           &  964  &  -5.33    &    0.077            \\
                                           &      {\bf present}                &   66                     &  5                          & 1142   & -5.02    &  0.079          \\
                                          &                                           &                        &  10                          & 975   & -5.33    &  0.076          \\\\
       Gliese 229 B    & Observations$^8$ &  71.4$\pm 0.6$      &  $<$10 &   &   -5.208$\pm 0.007$         &                            \\
                                      &     P2020                &   70.5                     &  8                   &  1171  &  -4.99    &    0.078            \\
                                         &                            &                              &     9                       & 1140   &   -5.04   &   0.077             \                 \\ 
                                        &                            &                             &      10                      & 1109   &   -5.09   &   0.077             \                 \\
                                            &      {\bf present}     &   70.5                     &  8                          & 1133   & -5.06    &  0.077          \\
                                        &                            &                            &   9                      & 1103  &   -5.11   &    0.076           \                 \\ 
                                        &                            &                            &   10                      & 1075 &   -5.16   &    0.076           \                 \\  
                                        &                           &   71.5                     &  10                          & 1110   & -5.10    &  0.076          \\
    \hline   \hline 
            \end{tabular}
         \end{center}
         \caption{Effective temperature, luminosity and radius of the various BDs examined in the text obtained with the P2020 and present models for the observed masses and ages. Observations: $^1$Cheetham et al. (2018), $^2$Chen et al. (2022), $^3$Bowler et al. (2018), $^4$Brandt et al. (2019), $^5$Cardoso (2012),  $^6$Dupuy et al. (2019), $^7$Sahlmann et al. (2020),  $^8$Brandt et al. (2021). Models are Phillips et al. (2020, P2020) and present calculations. We have taken 1 $\mjup$ = $9.5\times 10^{-4}\,\msol$ in the evolutionary models.}
\label{table_obs}
      \end{table*}

\subsection{HD4113C}

Using high-contrast imaging with the SPHERE instrument at the VLT, Cheetham et al. (2018) obtained the first images of the cold BD HD4113C (Sp=T9), which is part of a dynamical system with a M-dwarf companion. The dynamical mass is $\mdyn=66\pm5\,\mjup$, while comparison of the observed spectrum of HD4113C with atmospheric models (Morley et al. 2012, 2014 and Tremblin et al. 2015)  yields $\teff \sim500$-600 K, $\log\, g$=4.5-5.0 and a radius $R\sim 1.4$-1.5 $\rjup$, much larger than predicted for old, high-mass BDs. Using stellar evolution models (Mowlavi et al. 2012), the derived age of the parent star is found to be $\tau=5^{+1.3}_{-1.7}$ Gyr. The COND models (Baraffe et al. 2003) for such temperatures predict a mass $M=36\pm5\,\mjup$ for the BD, in strong conflict with the dynamical mass. Conversely, for $M=66\,\mjup$, these models predict $\teff\sim 1200\pm170$ K at the age of the system,  significantly higher than the afore estimates from the best-fit spectral models. Table \ref{table_obs} displays the temperature, luminosity and radius obtained with the Phillips et al. (2020) models, which use the CMS19 models, and the present ones, with the CD21 EOS. Both models use the same abundances for helium, $Y=0.275$, and metals, $Z_\odot=0.017$,
yielding an 'equivalent' helium abundance $Y_{eq}=0.292$.
This comparison highlights the impact of the H/He interactions in the most recent EOS models. As seen in the table, even though, for the relevant mass and age range, the models including the new EOS
yield temperatures cooler and luminosities fainter than the Baraffe et al. (2003) and Phillips et al. (2020) models, relieving part of the tension, it is clear that they are still far from resolving the disagreement with the observational determinations.

As noted by Cheetham et al. (2018), this discrepancy may be caused by the object being an unresolved binary brown dwarf system or by the presence of an additional object in the system, which could have biased the RV data and caused an overestimate of the dynamical mass.
An equal mass binary of 500-600 K objects with $R\sim 1\,\rjup$, notably, would provide a good match to the observed data while being in good agreement with the  model predictions,
namely $\teff=600\pm40$ K for a 33 $\mjup$ object at $5\pm1$ Gyr (e.g. Phillips et al. 2020 or present models). 

\subsection{Eps Indi Bab}

Recently, Chen et al. (2022) reported  dynamical masses for the binary BD system $\epsilon$ Indi Ba (Sp=T1-1.5) and  $\epsilon$ Indi Bb (Sp=T6, also called $\epsilon$ Indi C), with individual masses $\mdyn=66.92\pm0.36\,\mjup$ and  $\mdyn=53.25\pm0.29\,\mjup$, respectively,  with a $\approx 5\%$ precision. With an age of 3.5$^{+0.8}_{-1.0}$ Gyr
from $\epsilon$ Indi A's activity, this system provides a stringent constraint for BD cooling models, notably for old, massive BDs. 
Field-aged objects of spectral types T1 and T6 have effective temperatures in the range $\teff\simeq 1300$-900 K (e.g. Filipazzo et al. 2015). This is consistent with the inferred temperatures for Indi B and C, namely $\teff\simeq 1300$-1340 K and $\teff\simeq 880$-940 K, respectively (King et al. 2010).
Correct theoretical evolutionary models must thus allow these objects to reach these spectral types and temperatures within the aforementioned timescale for the observed metallicity of the host star $\epsilon$ Indi A ($[Fe/H]=0.13$).
None of the widely used  Chabrier et al. (2000), Burrows et al. (2001) and  Saumon \& Marley (2008) 'cloudy' models can fullfill this constraint and predict too high temperatures and luminosities. 
The  'COND' Baraffe et al. (2003) 
predict faster cooling rates, due to the less opaque atmospheres.
However, as noted in \S2,  the lower opacity in the COND atmosphere models stems, notably, from missing line opacities. Therefore, this analysis suggests that the aforementioned models underpredict 
the cooling rates for $\epsilon$ Indi Ba and Bb.

\begin{figure*}
\centering
{\includegraphics[height=9.5cm,width=9cm]{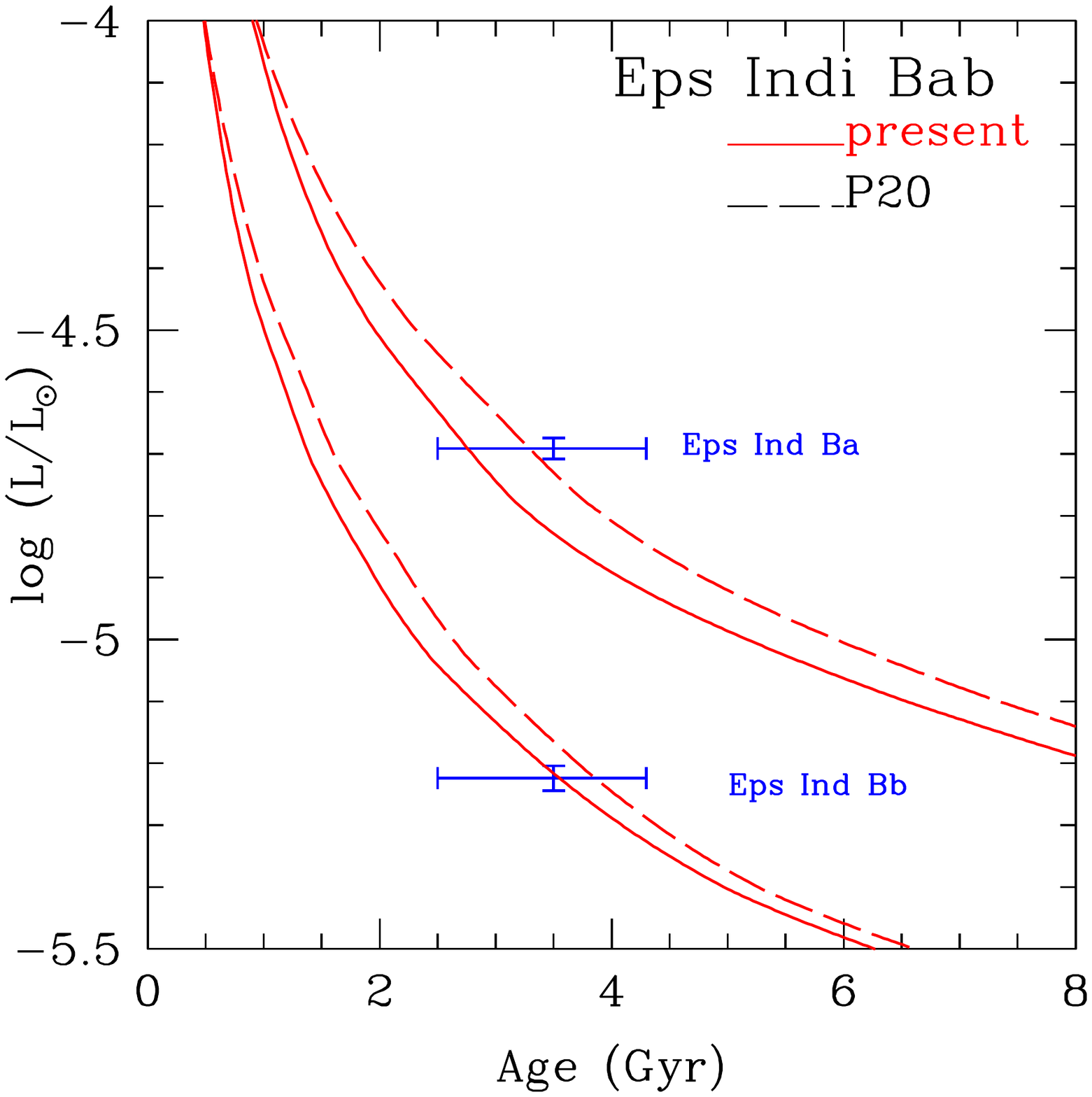}}  
{\includegraphics[height=9.5cm,width=9cm]{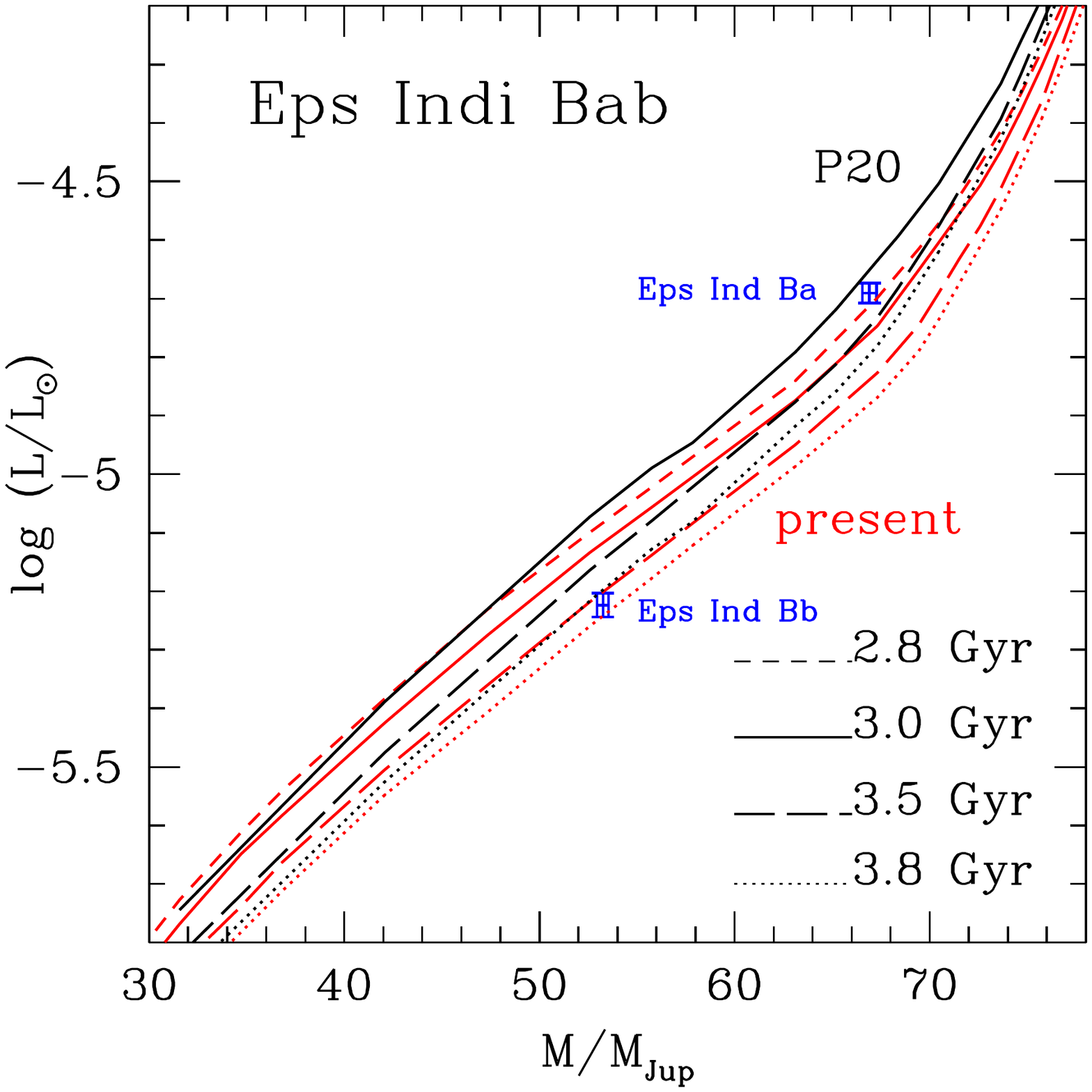}}  
\vspace{-1.5cm}
\caption{Left: cooling curves for a $67\,\mjup$ (top) and a $53\,\mjup$ (bottom) brown dwarf, respectively, representative of the $\epsilon$ Ind Bab system. Solid line: present models, based on the CD21 EOS; dashed line: Phillips et al. (2020) models, based on the CMS19 EOS.
Right: brown dwarf isochrones typical of the inferred age of the $\epsilon$ Ind Bab system, calculated with the present (red) and Phillips et al. (2020) (black) models.}
\label{Eps-Ind}
\end{figure*}

Figure \ref{Eps-Ind} displays the $L$-$t$ and $L$-$M$ relationships obtained for this system with the Phillips et al. (2020) cooling models, based on the CMS19 H/He EOS, and the present ones, based on the CD21 EOS, using the same ATMO atmosphere models and effective
helium+heavy element abundance $Y_{eff}=0.292$.
As seen in the Figure, while the Phillips et al. (2020) models yield an age of between $\sim 3.0$ and 3.8 Gyr for the system for the observed luminosities, the present model predict slightly younger ages, of 2.8 Gyr for Ba and 3.5 Gyr for Bb. Both models are in good agreement
with the observational determination. Assuming coevality, this corresponds to a $\sim 0.7$ Gyr uncertainty on the age of these BDs, for this age and mass range. As seen in Fig. 13 of Chen et al. (2022), the Saumon \& Marley (2008) hybrid models predict a significantly older age (5 Gyr) for
the system, at odds with the observational determination, suggesting a too slow cooling rate for these models. The corresponding effective temperatures and surface gravities between 2.8 and 3.5 Gyr with the present models are $\teff=$1354-1265 K,
$\log \,g=$5.45 for $\epsilon$ Indi Ba, for a mass 67 $\mjup$, and $\teff=$1060-998 K, $\log \,g=$5.34 for $\epsilon$ Indi Bb,  for a mass 53 $\mjup$, respectively (see Table 2). As illustrated in Figure \ref{Eps-Ind}  and discussed earlier, 
the faster cooling rate for the most massive BDs with the CD21 EOS than with the CMS19 one stems the sharp increase of degeneracy in this mass range (see e.g. Fig. 1 of Chabrier \& Baraffe 2000). The cooling rate then appears to be a bit too slow as a function of the mass, with $\Delta \log L/\Delta \log M\simeq 3.95$ instead of 5.37 between $\epsilon$ Indi Ba and Bb, i.e. from early T to late T, even though the discrepancy remains within 2$\sigma$. As mentioned earlier, this stems more likely from remaining issues with the atmosphere models than with the EOS.

\setlength{\unitlength}{1cm}

 \setlength{\unitlength}{1cm}
\begin{figure*}%
\centering
{\includegraphics[height=6.9cm,width=7.cm]{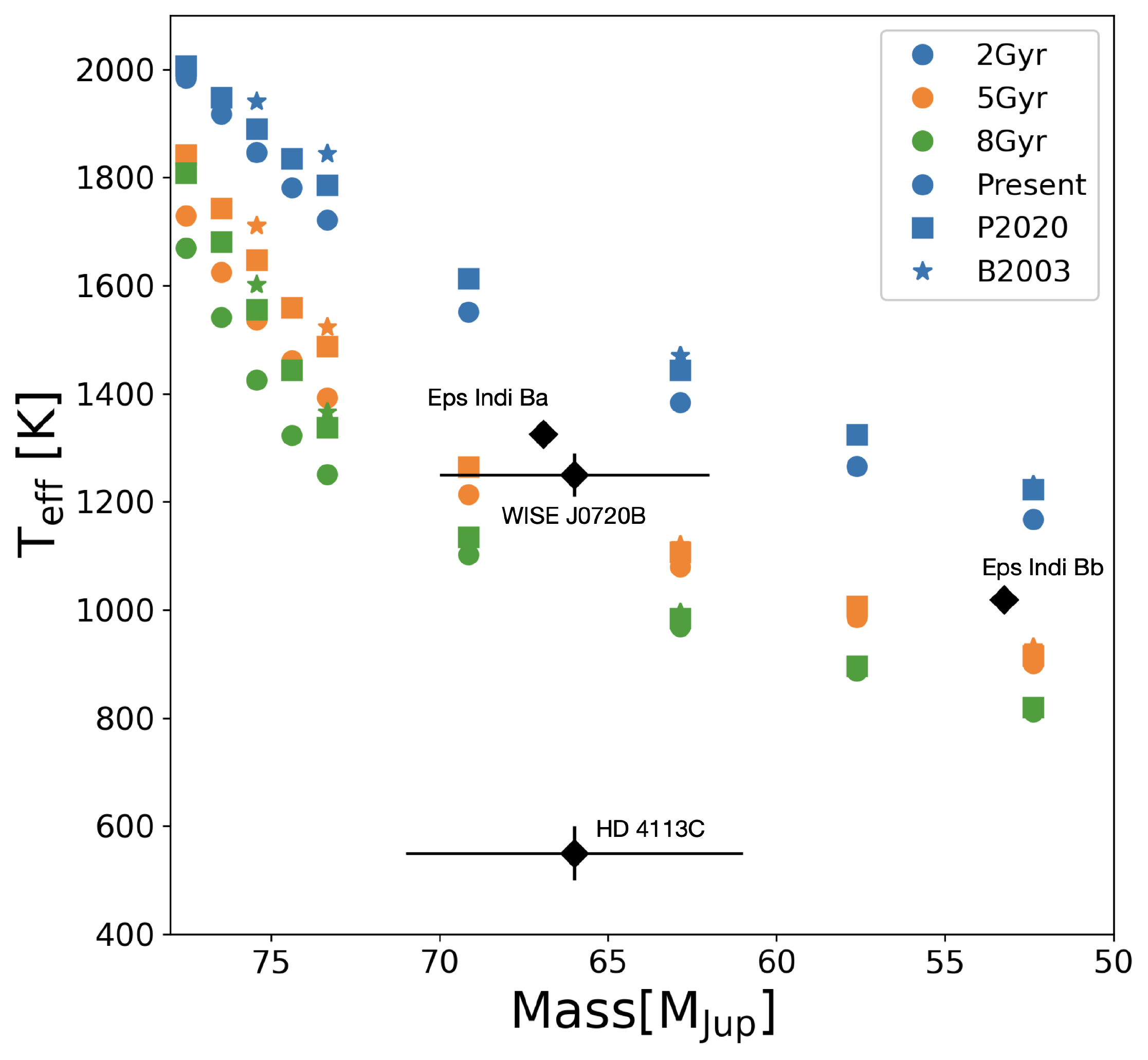} }
\hspace{1.cm}
{\includegraphics[height=6.9cm,width=8.5cm]{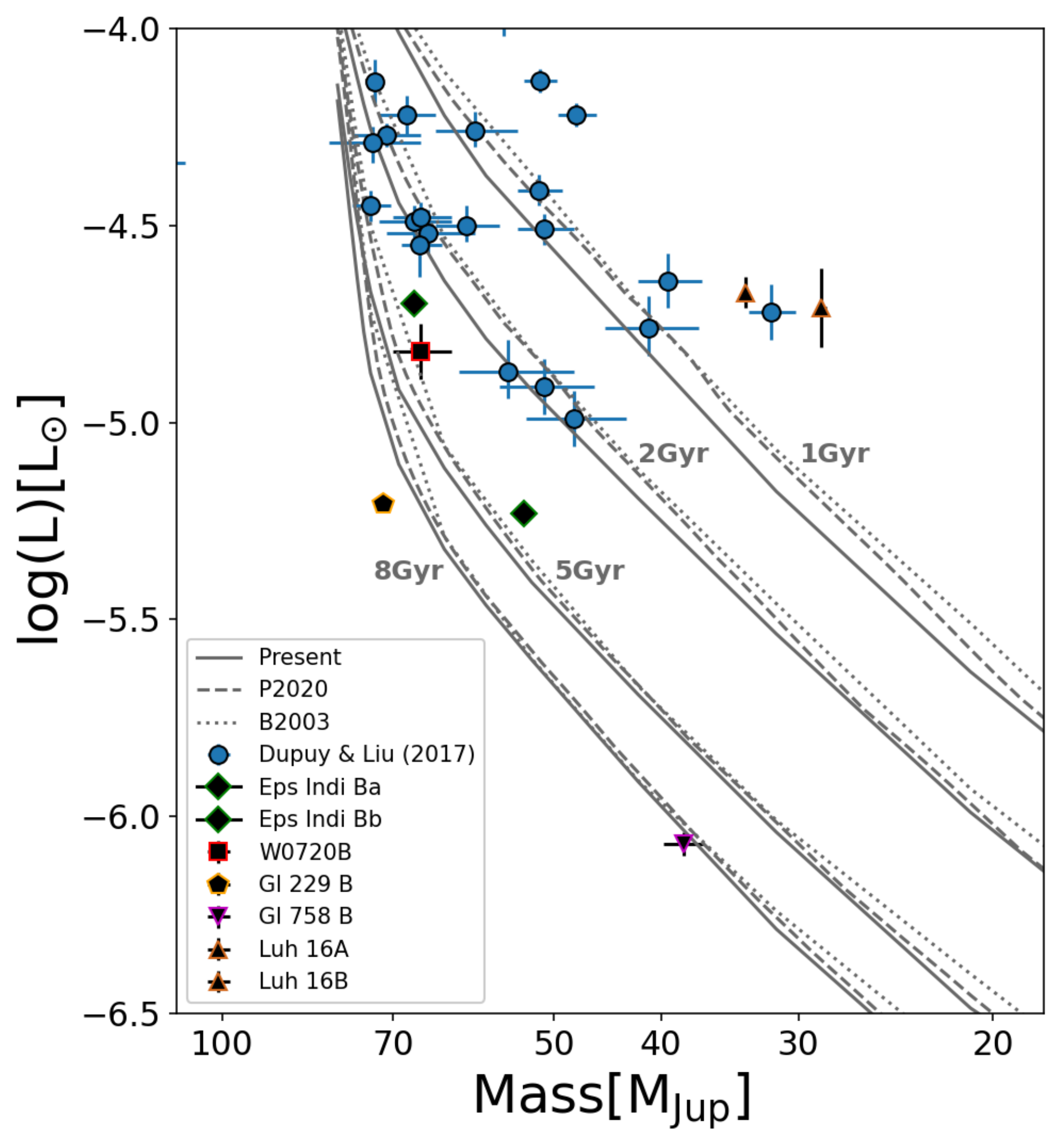} }
\caption{Left: effective temperature as a function of mass for massive BDs for 3 isochrones calculated with the present
models, based on the CD21 EOS and ATMO atmosphere models, the Phillips et al. (2020) models, based on the CMS19 
EOS and ATMO atmosphere models and the Baraffe et al. (2003) models, based on the SCvH EOS and the COND atmosphere m
odels. Right: luminosity as a function of mass for 3 isochrones for BDs with dynamical mass measurements. References
 for the  data are given in Table 2.
All models have an equivalent
helium mass fraction $Y_{eq} = 0.292$.}
\label{fig_data}
\end{figure*}

\subsection{Gl 758 B}

Combining radial velocity and astrometry, Bowler et al. (2018) determined a dynamical mass of $\mdyn=42^{+19}_{-7}\,\mjup$  for the T7-T8 BD Gl 758 B, with a robust lower limit of 30.5 $\mjup$ at the 4$\sigma$ level, for nominal ages 1-6 Gyr adopted for the host star (Vigan et al. 2016). More recently, using Hipparcos and Gaia data, Brandt et al. (2021)
derived the most precise mass measurement to date for this system, with $\mdyn=38.1^{+1.7}_{-1.5}\,\mjup$, while their analysis of activity and rotation of Gl 758A favors an age $\gtrsim$6 Gyr.
As for the previous objects, substellar evolutionary models generally underestimate the mass of Gl 758 B. As noted by the authors above, 
this discrepancy can be reconciled if the system is older, which is consistent with activity indicators and recent isochrone fitting of the host star, or alternatively if the models are systematically overluminous by $\approx$0.1-0.2 dex. Atmospheric model fitting yields for this object a bolometric luminosity $\log (L/L_\odot)=-6.07\pm0.03$, $\teff=650$ K and $\log g=5.0$. All current models essentially underpredict the mass for an age 1-6 Gyr or, alternatively, are overluminous by $>0.1$ dex at this age. This is in the opposite sense from results by Dupuy et al. (2009, 2014), who found that substellar cooling models underpredict the luminosities of brown dwarfs with dynamical masses by $\approx$0.2-0.4 dex. Altogether, the most likely explanation for the disagreement in mass probably resides in the age of Gl 758. Older ages of 6-9 Gyr would readily put the predicted and dynamical distributions in excellent agreement and are indeed suggested from the low activity level, lack of X-ray emission and slow projected rotational velocity (see Bowler et al. (2018) and references therein). Indeed, more recent isochrone fitting are converging on an older value that agrees better with activity indicators, with an average of $5.3$-$7.5$ Gyr (Brewer et al. 2016, Luck 2017). This is supported by the results displayed in Table \ref{table_obs}, with the new models yielding a nearly perfect agreement with
the observations for a mass $\sim 30$-$40\,\mjup$ and an age $\sim$5-8 Gyr.

\subsection{WISE system J072003.20}

Individual dynamical masses for the nearby M9.5+T5.5 binary  system WISE J0720-0846AB have been determined by Dupuy et al. (2019). The BD companion has a
mass $\mdyn=66\pm4\,\mjup$, an effective temperature $\teff= 1250\pm40$ K and a luminosity $\log (L/L_\odot)=-4.8\pm0.15$. This suggests an age older than a few Gyr,
consistent with the age estimates for the primary star. As shown in Table \ref{table_obs}, both the P2020 and present models yield  a nearly perfect agreement with the observational determinations for an age $\sim$3-4 Gyr, with the new ones predicting a slightly younger age (more rapid cooling).
For the primary star, WISE J0720-0846A, as noted in Dupuy et al. (2019), models (BHAC 2015) overestimate the luminosity for its mass ($\mdyn=99\pm6\,\mjup$), or conversely underestimate its mass for its luminosity,  at about 2$\sigma$  (see their Table 5 and Fig. 7). As mentioned earlier, the new CD21 EOS does not significantly change this analysis for such high (stellar) masses (see Table 1).

\noindent    It is worth  noting that the mass and age determinations of WISE J0720-0846B are very similar to the ones of the BD HD4747B ($66\pm 3\,\mjup$, $2.9^{+0.5}_{-0.4}$ Gyr, $\log (L/L_\odot)=-4.55\pm0.08$; Brandt et al. 2021, Table 10), for which the present models give a very good agreement (see Table 2). 

\subsection{2M1059 and 2M0805}

Sahlmann et al. (2020) have measured the complete astrometric orbits for the systems 2M0805+48 and 2M1059-21. They find masses
$\mdyn=66^{+5}_{-14}\,\mjup$ and spectral type T5.5 for 2M0805+48 B and $\mdyn=67^{+4}_{-5}\,\mjup$, T3.5  for 2M1059-21 B.

The striking feature of their analysis is that the mass for the T3.5 2M1059-21B object is significantly higher than its two spectral type equivalents DENIS J2252-1730B (T3.5), 41$\pm 4\, \mjup$, and 2MASS J1534-2952A (T4.5), 51$\pm 5\, \mjup$.
As noted by the authors, the mass of 2M0805+48B is almost equal to the one of WI0720-08B estimated by Dupuy et al. (2019), which has the same spectral type, yet a mass higher than the other three T5 dwarfs with measured dynamical masses, which all have masses lower than $<61\,\mjup$ as an upper limit (see Table 6 of Dupuy et al. 2019 ).
As shown by the authors (see their Fig. 13), the masses derived for each member of our two pairs are compatible  at the 1$\sigma$ level with the 5 to 12 Gyr isochrones 
of Baraffe et al. (2015), while $>$1 Gyr isochrones show reasonable agreement with the observational data.

This, as the other previous analysis, highlights the fact that the higher observational masses than other T-dwarfs of similar spectral type and than predicted by the models,
suggesting that models underpredict the mass for a given temperature or luminosity, concerns essentially the most massive ($\gtrsim 60\,\mjup$) BDs, old enough to have reach the T spectral type domain, i.e. older than $\gtrsim 1$ Gyr for such masses. This strengthens our
suggestion that this property stems essentially from the higher degeneracy, thus the faster cooling of these objects, which culminates at the highest central density of the stellar-substellar domain (see Fig. 1 of Chabrier \& Baraffe 2000).
The new models predict cooler temperatures and fainter luminosities for a given mass and age compared with the Baraffe et al. (2003) models, 
by up to about 100-150 K and $\sim 0.2$-$0.3$ dex for the present ones, resolving at least partly the discrepancies mentioned above (see Table \ref{table_obs}).

\noindent   Here again, we note that the mass and age of this system are  similar to the ones of the BD HD19467B listed in  Brandt et al. (2021), with an excellent agreement for the models. 
    
\subsection{Gliese 229B}

Combining  Keck/HIRES radial velocities, imaging with HiCIAO/Subaru and the HST, and absolute astrometry from Hipparcos and GAIA, Brandt et al. (2020) measured the dynamical mass $\mdyn=70\pm 5\,\mjup$  for the T7 BD Gliese 229B.
Not only this value is higher than the $\simeq 64.0^{+2}_{-1}\,\mjup$ predicted by the Baraffe et al. (2003) or Saumon \& Marley (2008) models for such a low luminosity, $\log(L/L_\odot)=-5.208$, but to be compatible with the observational determinations,
the models would predict an age of 7-10 Gyr. Such an age seems to be excluded by kinematic and activity indicators that rather suggest a range 2-6 Gyr. Gliese 229B thus joins the  club of ultracool BDs near the HBMM too massive, for their age,
for the  model predictions.
For a $70\,\mjup$ at an age of 10 Gyr, the Phillips et al. (2020) models (see their Fig. 8), calculated with the Chabrier et al. (2019) EOS, are $\sim 0.1$ dex less luminous than the Baraffe et al. (2003, B03) and $\sim 0.4.$ dex less luminous than the hybrid cloud tracks of SM08,
helping to relieve some of the tension. As seen in Table \ref{table_obs}, the present models, based on the new CD21 EOS, predict $\gtrsim$40 K cooler and $\sim -0.1$ dex fainter models than the Phillips et al. (2020) ones for a mass and luminosity consistent with the latest observationally inferred values, for an age of about 9-10 Gyr, improving the model-observations agreement. 

Best-fit atmosphere models to the observed spectrum of Gliese 229B by Nakajima et al. (2015) yield acceptable solutions in the range 750 K$\le\teff\le$ 900 K, $4.75\le \log g\le 5.0$. As seen in the Table, these values are quite small compared to the ones from the models.
It must be stressed, however, that this fitting procedure is based on one single source of {\it cloudy} atmosphere models, namely Tsuji (2002, 2005). A more robust determination of the effective temperature and gravity requires further detailed comparisons with more
recent atmosphere models, a point we have already stressed in \S2. 

\noindent As a last remark, it is worth noting that a possibility to resolve the discrepancy between models and observations for Gl229 B would be that this
latter is itself an unresolved tight binary (see e.g. Brandt et al. 2021).

A case close to Gliese 229B is HR 7672 B, with a dynamical mass $\mdyn=72.7\pm{0.8}\,\mjup$ and a weighted average luminosity $\log(L/L_\odot)=-4.25\pm 0.05$ (Brandt et al. 2021), while the
activity analysis suggests a rather young age, centered around 2 Gyr (Brandt et al. 2019, 2021). For this age and a mass $M=70\,\msol$ ($=72.7\,\mjup$), the present models predict $\log(L/L_\odot)=-4.25$, $\teff=$1730 K, while
the Phillips et al. (2020) ones yield $\log(L/L_\odot)=-4.16$, $\teff=$1785 K. 
\bigskip

In order to illustrate the results listed in Table 2, Fig. \ref{fig_data} displays the $\teff$-mass and $L$-mass data comparisons (as in Fig. 8 of Phillips et al. 2020), including the most recent determinations, for massive BDs for 4 isochrones, namely 1, 2, 5 and 8 Gyr, based on the CMS19 (Phillips et al. 2020) and CD21 (present) EOS models.

\bigskip

 To briefly summarise this section, we note that the models based on the most recent EOSs, notably the latest CD21 one, reach cooler temperatures and fainter luminosities than the previous generation for a given mass and age, illustrating the faster cooling rates for massive, old ($\gtrsim 1$ Gyr) BDs, yielding in general a much better agreement with the observational determination. As seen in Table  \ref{table_obs}, some tension remains potentially for Gl229B but most importantly for HD4113C, which appears to be substantially cooler and fainter than predicted by the models. Given the general agreement for all the other objects, this suggests that these two BDs could be  unresolved systems.

\section{Conclusion} 

In this paper, we have explored the impact of the latest EOS for dense hydrogen-helium mixtures upon the structure and evolution of very low mass stars and brown dwarfs.
Whereas the previously used Saumon et al. (1995) and Chabrier et al. (2019) EOSs are based on the so-called additive volume law, which implies that the interactions between hydrogen and
helium species are not taken into account, these interactions are included in the Chabrier \& Debras (2021) EOS, based on the QMD simulations of Militzer \& Hubbard (2013) (see CD21 for details). These interactions modify the thermodynamic properties of the H/He mixture in two ways. They yield cooler and denser, thus more compact structures, i.e. smaller radii and thus fainter luminosity for given mass and age. They also yield cooler entropy profiles, a quantity of prime importance for these fully convective bodies, which affects the onset and the development of electron degeneracy throughout the body (see Fig. 9 of CD21).  This translates into a faster cooling rate and thus larger masses for given effective temperatures and luminosities at a given age in the brown dwarf regime (see Table \ref{table_obs}). 
Confronting these new models with several observationally determined BD dynamical masses, we show that, indeed, this improves the agreement between evolutionary models and observations and resolves at least part of the 
observed discrepancies for massive, rather old brown dwarfs. In the stellar domain, the impact of the new EOS is inconsequential.
Resolving the remaining discrepancies between BD observations and models probably requires further improvements of atmosphere models and self-consistent evolutionary calculations with these models. Work in this direction in under progress.

A noticeable consequence of this improvement in the dense H/He EOS is that it yields a larger H-burning minimum mass, now found to be $0.075\,\msol$  ($\sim78\,\mjup$) with the ATMO atmosphere models for
solar helium and heavy element abundances, $Y=0.275$, $Z_\odot=0.017$, respectively,  thus an equivalent helium abundance $Y_{eq}\simeq Y+Z=0.292$.
The corresponding radius, effective temperature and luminosity at 10 Gyr are listed in Table 1.

These new brown dwarf models will be available on the sites http://perso.ens-lyon.fr/isabelle.baraffe/CBPD2022 and http://opendata.erc-atmo.eu


\begin{acknowledgements}
The authors are grateful to the referee, Timothy Brandt, whom remarks helped clarifying the manuscript. This work has been supported by the consolidated STFC Consolidated Grant ST/V000721/1, the ERC grant No. 787361-COBOM and the Programme National de Plan\'etologie (PNP).
\end{acknowledgements}

\end{document}